\documentclass[final,1p,times]{elsarticle} % do not change
\usepackage{graphicx} % do not change
\usepackage{amssymb} % do not change
\usepackage{amsthm} % do not change
\usepackage{lineno} % do not change

\usepackage{url}   % added by author

\journal{Nuclear Physics A} % do not change
\begin{document} % do not change

\begin{frontmatter} % do not change

%% QM09Author: please enter your  
%% Title, author and address info here; please do not use footnotes

% Your Title - please insert
\title{Quark matter in neutron stars}

% Principle author, and co-authors - please insert
\author{Mark G. Alford}

% Address - please insert
\address{Physics Department \\ Washington University CB 1105 \\
Saint Louis, MO 63130 \\ USA}

\begin{abstract} % do not change
According to quantum chromodynamics,
matter at ultra-high density and low temperature is
a quark liquid, with a condensate of Cooper pairs of quarks 
near the Fermi surface (``color superconductivity'').
This paper reviews the physics of color superconductivity,
and discusses some of the proposed signatures by which we might
detect quark matter in neutron stars.
\end{abstract} % do not change

\end{frontmatter} % do not change

%% QM09: we keep linenumbers at least for initial version
% \linenumbers % do not change

%% start of main text - please insert. 

\newcommand{\Dash}{\boldmath $-$}
\newcommand{\MeV}{{\rm MeV}}
\newcommand{\fm}{{\rm fm}}
\newcommand\eqn[1]{(\ref{#1})}      % parentheses around the LaTex "ref" macro
\newcommand\Eqn[1]{Eq.~(\ref{#1})}  % includes ``Eq.'' in front

\newcommand{\al}{\alpha}
\newcommand{\be}{\beta}
\newcommand{\ga}{\gamma}
\newcommand{\Ga}{\Gamma}
\newcommand{\de}{\delta}
\newcommand{\De}{\Delta}
\newcommand{\ep}{\varepsilon}
\newcommand{\eps}{\epsilon}
\newcommand{\ze}{\zeta}
\newcommand{\ka}{\kappa}
\newcommand{\la}{\lambda}
\newcommand{\La}{\Lambda}
\newcommand{\ph}{\varphi}
\newcommand{\del}{\nabla}
\newcommand{\si}{\sigma}
\newcommand{\Si}{\Sigma}
\renewcommand{\th}{\theta}   % LaTeX: \th already defined
\newcommand{\Up}{\Upsilon}
\newcommand{\om}{\omega}
\newcommand{\Om}{\Omega}

\section{Introduction}
\label{sec:intro}
One of the most striking features of quantum chromodynamics
(QCD) is asymptotic freedom: the
force between quarks becomes arbitrarily weak as the characteristic
momentum scale of their interaction grows larger.  This immediately
suggests that at sufficiently high densities and low temperatures,
matter will consist of a Fermi sea of essentially free quarks,
whose behavior is dominated by  the
high-momentum quarks that live at the Fermi surface.

However, over the last decade it has become clear
that the phase diagram of dense matter is much richer than this.
In addition to the hadronic phase with which we are
familiar and the quark gluon plasma (QGP) that 
exists at temperatures above about $170~{\rm MeV}$, there is a
whole family of ``color superconducting'' phases, which are expected
to occur at high density and low temperature;
for a detailed review, see Ref.~\cite{Alford:2007xm}.
These phases have observational importance, because
they may occur naturally in the
universe, in the cold dense cores of compact (``neutron'') stars,
where densities are above nuclear density, and temperatures are of the
order of tens of {\rm keV}.  It might conceivably be possible to
create them in future low-energy heavy ion colliders, such as 
% the Japan Proton Accelerator Research Complex (J-PARC) or 
the Compressed Baryonic Matter facility at GSI Darmstadt \cite{CBM}.
Up to now, most work on
signatures has focussed on properties of color superconducting quark
matter that would affect observable features of compact stars, and we
will discuss some of these below.

% \clearpage

\section{Color superconductivity}

\subsection{Cooper pairing of quarks}
The essential physics of color superconductivity is the same
as that underlying conventional superconductivity in metals,
and also superfluidity in liquid Helium, nuclear matter, and cold
atomic gases. The crucial ingredients are
a Fermi surface and an attractive interaction between the
fermions. Quark matter has exactly these ingredients.
It was shown by Bardeen, Cooper, and
Schrieffer (BCS) \cite{BCS} that if there is {\em any} channel in which the
fermion-fermion interaction is attractive, then there is a state
of lower free energy than a simple Fermi surface. That state arises
from  a complicated coherent 
superposition of pairs of particles (and holes)---``Cooper pairs''.
This can easily be understood
in an intuitive way.
The free energy at zero temperature is $F= E-\mu N$, where $E$ is
the total energy of the system, $\mu$ is the chemical potential for
fermion number, and
$N$ is the number of fermions. The Fermi surface is defined by a
Fermi energy $E_F=\mu$, at which the free energy is minimized, so
adding or subtracting a single particle costs zero free energy. 
Now switch on a weak attractive interaction.
As we have just seen, it costs negligible free energy to
add a pair of particles (or holes) close to the Fermi surface, 
and if they have the
right quantum numbers then the attractive
interaction between them will lower the free energy of the system.
Many such pairs will therefore
be created in the modes near the Fermi surface, and these pairs,
being bosonic, will form a condensate. The ground state will be a
superposition of states with all numbers of pairs, 
spontaneously breaking the fermion number symmetry. 

High-density low-temperature quark matter has exactly the right 
ingredients for the BCS mechanism to operate.
Asymptotic freedom of QCD means that
at sufficiently high density and low temperature,
there is a  Fermi surface of almost free quarks. 
And the interactions between quarks near the Fermi surface are 
certainly attractive in some channels, because quarks bind
together to form baryons.
We therefore expect quark matter that is sufficiently cold and dense
to {\em generically} exhibit color superconductivity.
The densities at which the strong interaction becomes
perturbatively weak are extraordinarily high \cite{Rajagopal:2000rs},
so it remains an open question whether color superconducting phases
persist down to the densities achieved in neutron star cores.

The phase structure of cold quark matter is expected to be
complicated, with many competing phases (see Fig.~\ref{fig:phase}). 
This is because
quarks, unlike electrons, have color and flavor as well as spin
degrees of freedom, so many different patterns of pairing are possible.
Since pairs of quarks cannot be color singlets,
the resulting condensate will break the local color symmetry
$SU(3)_{\rm color}$.  We therefore call Cooper pairing
of quarks ``color superconductivity''.
Note that the quark pairs play the same role here as the Higgs particle
does in the standard model: the color-superconducting phase
can be thought of as the Higgs phase of QCD.

The wavefunction of a Cooper pair must be antisymmetric under exchange
of the two fermions. The most attractive channel for two quarks is 
color antisymmetric (the color $\bar{\bf 3}_A$), Dirac antisymmetric
(the Lorentz scalar $C \gamma_5$), and spatially symmetric ($s$-wave).
This requires antisymmetry in the remaining label, flavor.
We conclude that pairing between different flavors will be 
typically be the energetically favored option. As we will see,
this turns out to
be crucial to understanding the high-density phase structure of quark matter.

\subsection{Phase diagram of quark matter}

\begin{figure}[t]
\includegraphics[width=0.5\hsize]{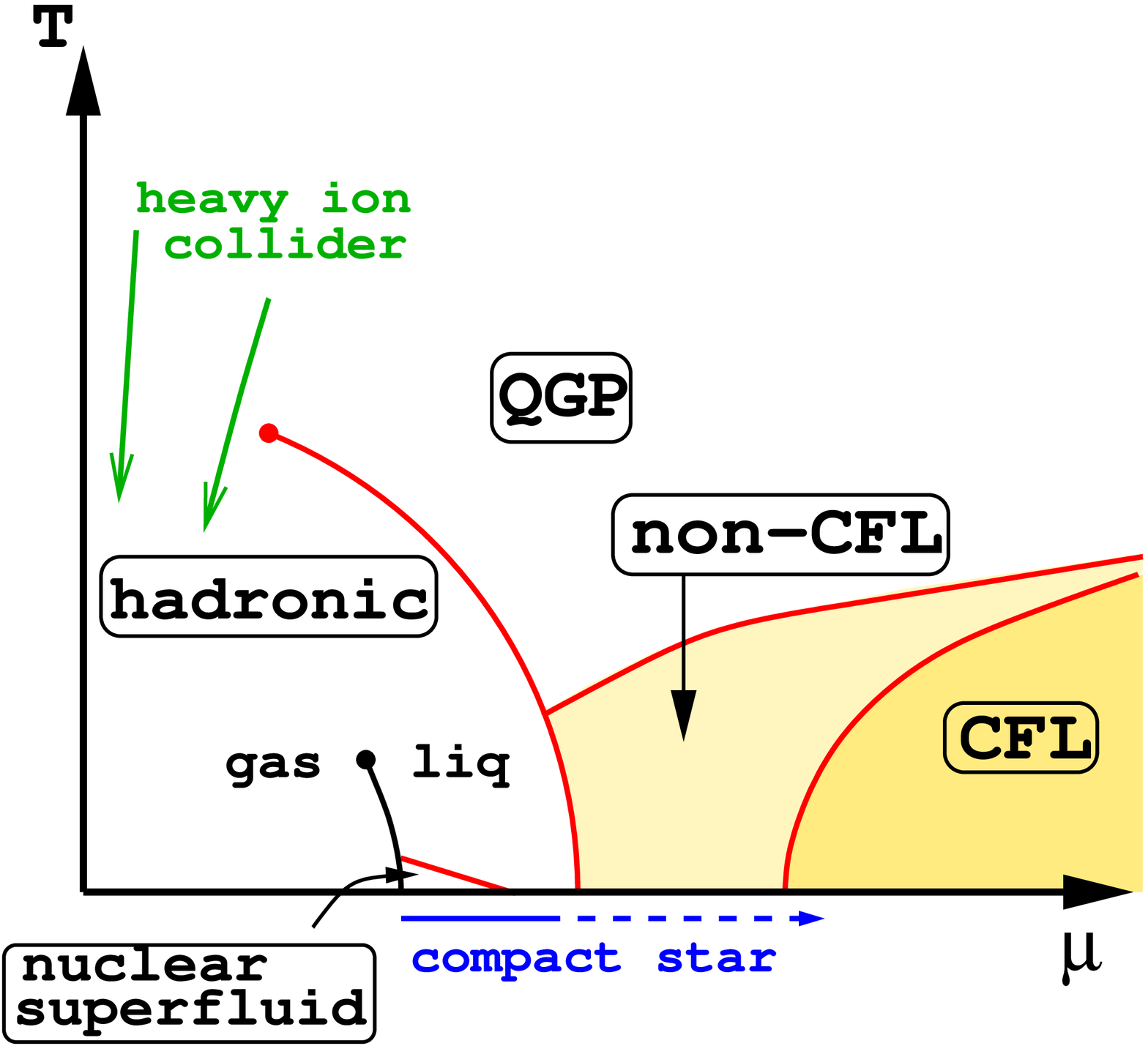}%
\includegraphics[width=0.5\hsize]{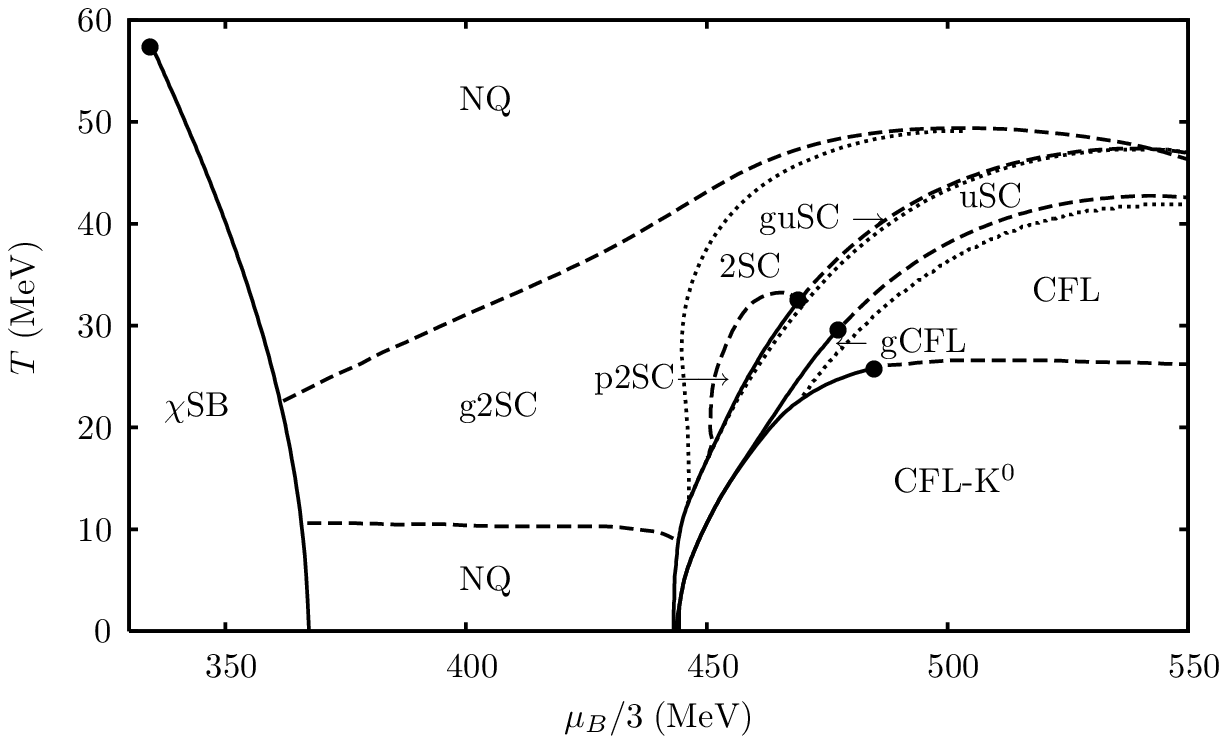}
\caption{Left panel: conjectured phase diagram of matter
at ultra-high density and temperature. The CFL phase is a 
color superconductor that is superfluid
(like cold nuclear matter) and has broken chiral symmetry (like the
hadronic phase).
Right panel: calculated phase diagram using an NJL model
and considering only spatially uniform phases \cite{Warringa:2006dk}.
}
\label{fig:phase}
\end{figure}

Fig.~\ref{fig:phase} (left panel) 
shows a schematic phase diagram for QCD that is
consistent with what is currently known.
Along the horizontal axis the temperature is zero, and the 
baryon density is zero
up to the onset transition where it jumps to nuclear density; the density 
then rises with increasing $\mu$.
Neutron stars are in this region of the phase diagram, although it is 
not known whether their cores are dense enough to reach the quark matter 
phase. Along the vertical axis the temperature rises, taking us through 
the crossover from a hadronic gas to the quark-gluon plasma. This is the 
regime explored by high-energy heavy-ion colliders.

At the highest densities we find the color-flavor locked (CFL)
color-superconducting phase, in which the strange quark participates 
symmetrically with the up and down quarks in Cooper 
pairing. The CFL phase may extend all the way down to a few times
nuclear density, or there may, as shown in the figure, be
an interval of some other phase or phases.
These may include two-flavor color superconductivity (``2SC''),
crystalline color superconductivity (``LOFF'') 
\cite{Alford:2000ze,Casalbuoni:2003wh} or
% ,Rajagopal:2006ig
some form of single-flavor pairing 
\cite{Iwasaki:1994ij,Iwasaki:1995uw,Schafer:2000tw,Alford:2002kj,Buballa:2002wy}.
% ,Alford:2005yy}.
%The strange quark plays a crucial role in the phases of QCD,
%and we expect it to remain fully paired with the light flavors as
%long as $\mu \gg M_s^2/\De$, where $\De$ is a gap parameter for
%the pairing of the strange quark.

% In both cases, along the horizontal axis the temperature is zero, and
% the density rises from the onset of nuclear matter through the
% transition to
% quark matter. Compact stars are in this region of the phase diagram,
% although it is not known whether their cores are dense enough
% to reach the quark matter phase.
% Along the vertical axis the temperature rises, taking
% us through the crossover from a hadronic gas to the quark gluon plasma.
% This is the regime explored by high-energy heavy-ion colliders.

\subsection{Color-flavor locking (CFL)}
\label{sec:CFL}

At the highest densities, the favored pairing pattern is
``color-flavor locking'' (CFL) \cite{Alford:1998mk}.  This has been
confirmed by weak-coupling QCD calculations 
\cite{Schafer:1999fe,Shovkovy:1999mr}, Dyson-Schwinger calculations
\cite{Nickel:2006vf}, and in Nambu--Jona-Lasinio 
models \cite{Alford:1998mk,Schafer:1999pb,Evans:1999at}.
The CFL pairing pattern is
\begin{equation}
\begin{array}{c}
\langle q^\alpha_i q^\beta_j \rangle^{\phantom\dagger}_{1PI}
\propto C \gamma_5 \Bigl(
 (\kappa+1)\delta^\alpha_i\delta^\beta_j + (\kappa-1) \delta^\alpha_j\delta^\beta_i \Bigr)
% \langle q^\alpha_{i} q^\beta_j \rangle \sim \delta^\alpha_i\delta^\beta_j + \kappa\, \delta^\alpha_j\delta^\beta_i
  \\[2ex]
 {[SU(3)_{\rm color}]}
 \times \underbrace{SU(3)_L \times SU(3)_R}_{\displaystyle\supset [U(1)_Q]}
 \times U(1)_B 
 \to \underbrace{SU(3)_{C+L+R}}_{\displaystyle\supset [U(1)_{{\tilde Q} }]} 
  \times \mathbb{Z}_2
\end{array}
\label{CFLcond}
\end{equation}
Color indices $\alpha,\beta$ and flavor indices $i,j$ run from 1 to 3,
Dirac indices are suppressed,
and $C$ is the Dirac charge-conjugation matrix.
The term multiplied by $\kappa$ corresponds to pairing in the
$({\bf 6}_S,{\bf 6}_S)$, which
although not energetically favored
breaks no additional symmetries and so
$\kappa$ is in general small but not zero 
\cite{Alford:1998mk,Schafer:1999fe,Shovkovy:1999mr,Pisarski:1999cn}.
The Kronecker deltas connect
color indices with flavor indices, so that the condensate is not
invariant under color rotations, nor under flavor rotations,
but only under simultaneous, equal and opposite, color and flavor
rotations. Since color is only a vector symmetry, this
condensate is only invariant under vector flavor+color rotations, and
breaks chiral symmetry. The features of the CFL pattern of condensation are
\begin{itemize}
\setlength{\itemsep}{-0.7\parsep}
\item[\Dash] The color gauge group is completely broken. All eight gluons
become massive. This ensures that there are no infrared divergences
associated with gluon propagators, so at asymptotically high densities
this phase can be rigorously studied in perturbation theory.
\item[\Dash]
All the quark modes are gapped. The nine quasiquarks 
(three colors times three flavors) fall into an ${\bf 8} \oplus {\bf 1}$
of the unbroken global $SU(3)$, so there are two
gap parameters. The singlet has a larger gap than the octet.
\item[\Dash] 
A ``rotated electromagnetism''
survives unbroken. Its generator is ${\tilde Q}$, a linear combination of
a color rotation and an electromagnetic phase rotation; 
its gauge boson is therefore a combination
of the original photon and one of the gluons.
The CFL phase is electrically neutral without any electrons 
\cite{Rajagopal:2000ff}, and is therefore a transparent insulator.
\item[\Dash] Two global symmetries are broken,
the chiral symmetry and baryon number, so there are two 
gauge-invariant order parameters
that distinguish the CFL phase from the QGP,
and corresponding Goldstone bosons which are long-wavelength
disturbances of the order parameter. 
When the light quark mass is non-zero it explicitly breaks
the chiral symmetry and gives a mass
to the chiral Goldstone octet, but the CFL phase is still
a superfluid, distinguished by its baryon number breaking.
\item[\Dash]
The symmetries of the
3-flavor CFL phase are the same as those one might expect for 3-flavor
hypernuclear matter \cite{Schafer:1999pb,Alford:1999pa}, 
so it is possible that there is no phase transition between them.
\end{itemize}

\section{Cooper pairing in the real world: 2+1 flavors}

\begin{figure}
\parbox[t]{0.33\hsize}{
\begin{center} Unpaired\end{center}
\includegraphics[width=\hsize]{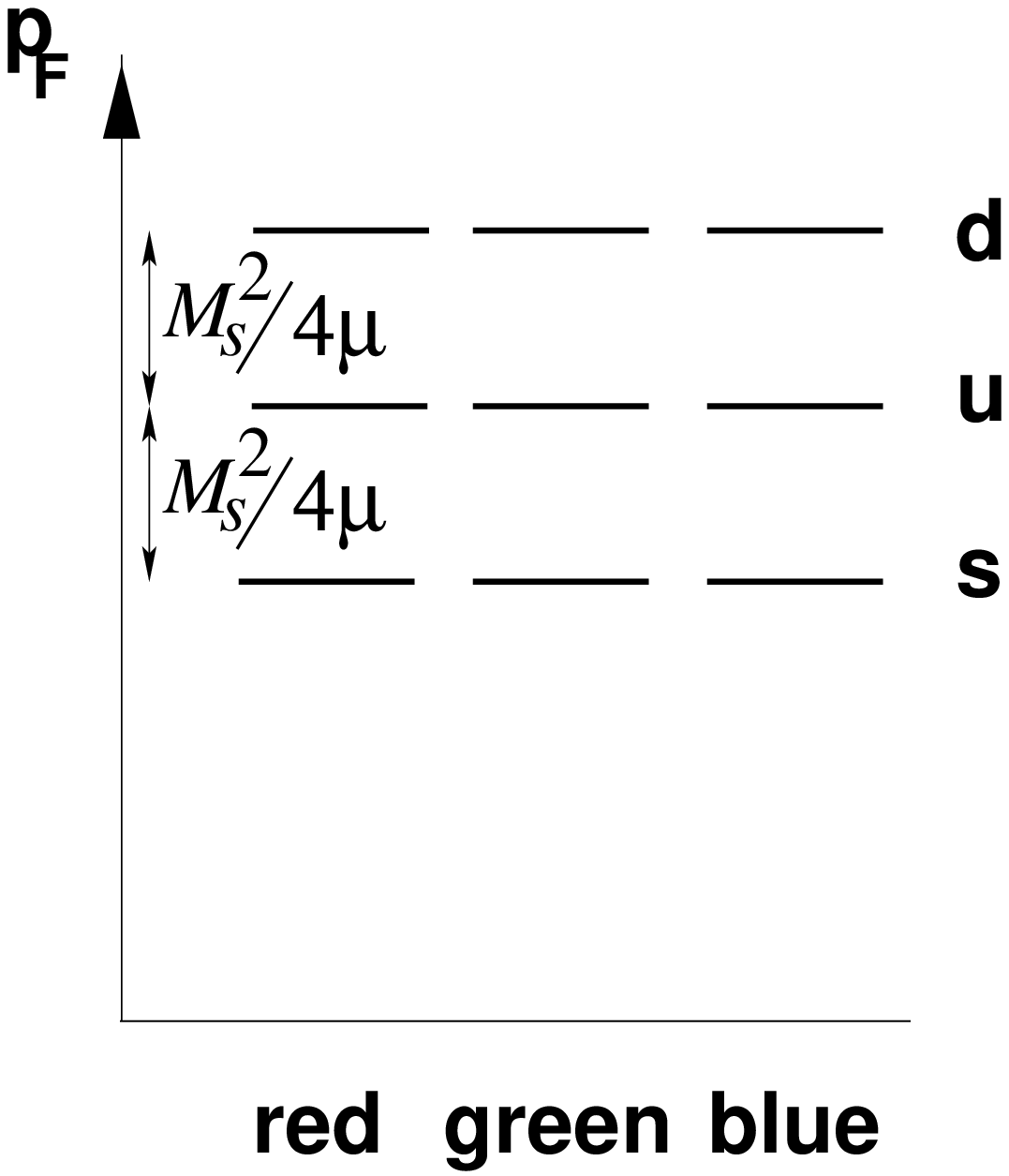}
}\parbox[t]{0.33\hsize}{
\begin{center} 2SC pairing\end{center}
\includegraphics[width=\hsize]{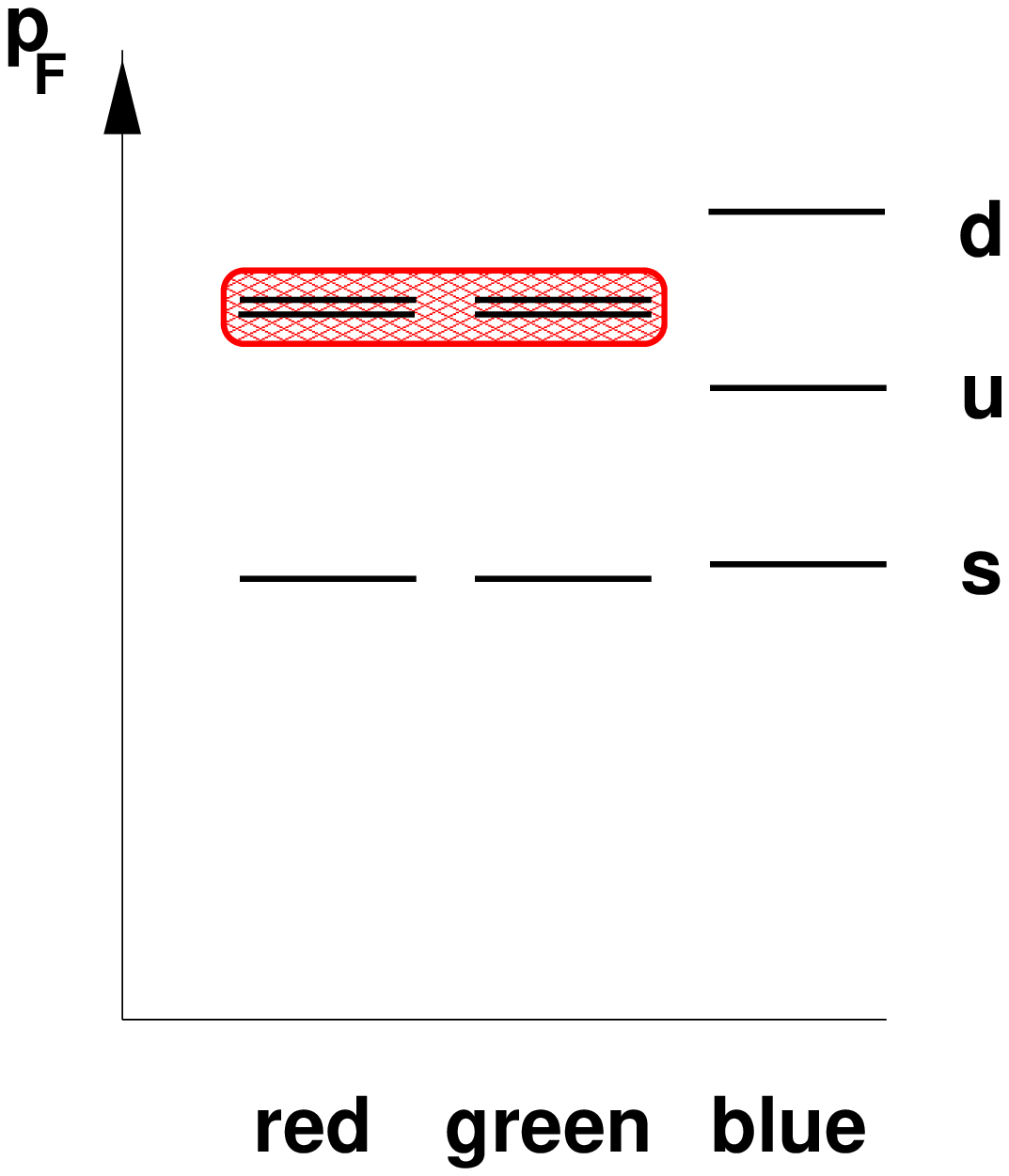}
}\parbox[t]{0.33\hsize}{
\begin{center} CFL pairing\end{center}
\includegraphics[width=\hsize]{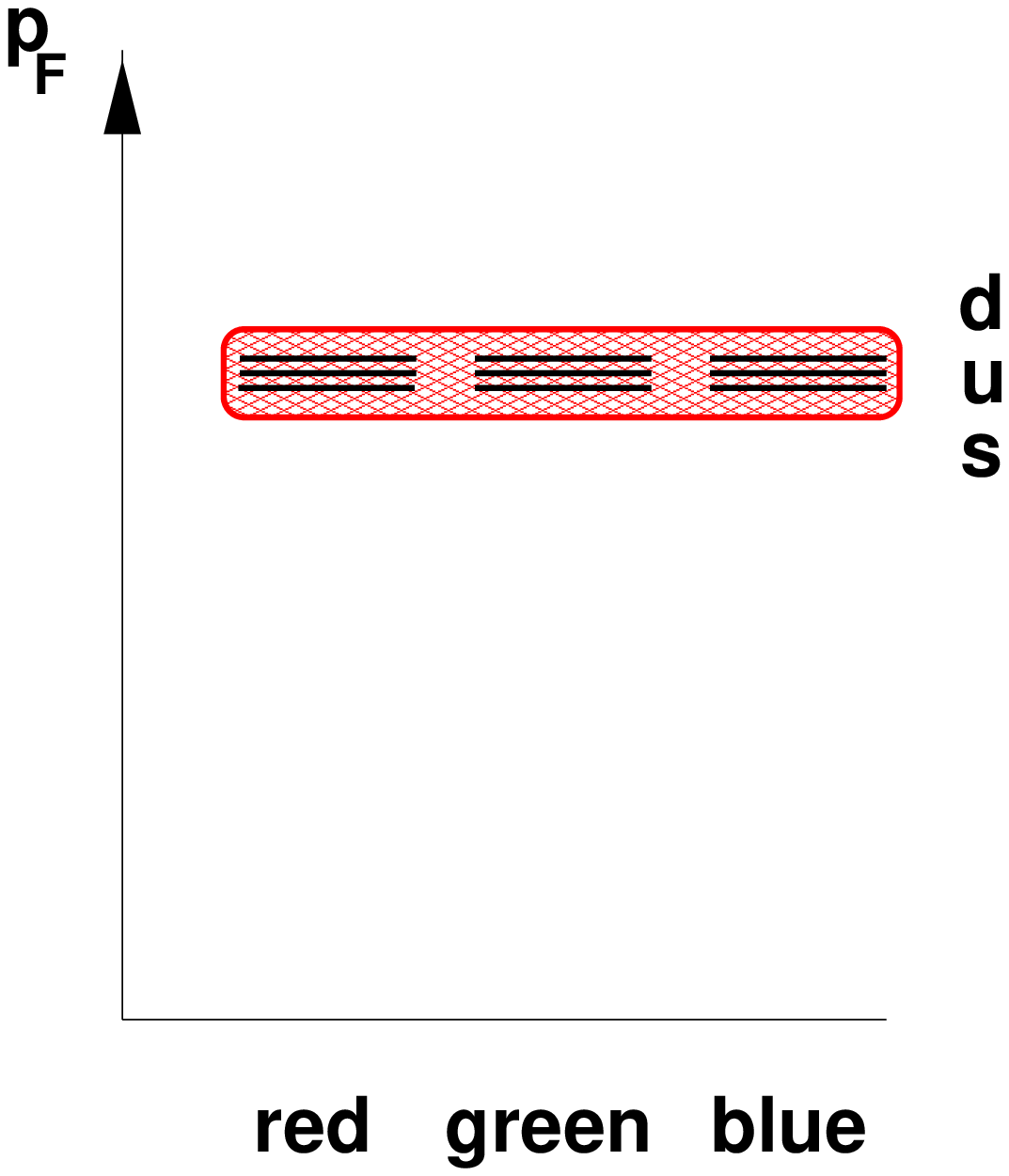}
%\begin{center} No electrons! \end{center}
}
\caption{(Color online)
Illustration of the splitting apart of the Fermi momenta of the
various colors and flavors of quarks (exaggerated for easy
visibility). In the unpaired phase,
requirements of neutrality and weak interaction equilibration
cause separation of the Fermi momenta of the various flavors.
The splittings increase with decreasing density, as $\mu$
decreases and $M_s(\mu)$ increases. At very high density
the splitting is small, favoring the the CFL phase, where
all colors and flavors pair and have a common Fermi momentum.
At intermediate density we expect complicated compromises between
pairing and Fermi-surface splitting, for example the 2SC phase, 
where up and down quarks of two colors pair,
locking their Fermi momenta together. 
}
\label{fig:splitting}
\end{figure}

In a real compact star we must require electromagnetic and color
neutrality \cite{Iida:2000ha,Alford:2002kj,Gerhold:2003js,Buballa:2005bv},
allow for equilibration under the weak interaction, and include a
realistic mass for the strange quark.  These factors tend to pull
apart the Fermi momenta of the different quark species, imposing an
energy cost on cross-species pairing, and hence disfavoring the CFL phase
at sufficiently low densities.
To see how this occurs, consider the left panel of Fig.~\ref{fig:splitting},
which shows the Fermi momenta of the different colors and flavors of the
quark species. The strange quarks have a lower Fermi momentum because 
they are heavier, and hence are more energetically costly.
To maintain electrical neutrality, the number of down quarks must be
correspondingly increased, so the down quark Fermi momentum is raised.
To lowest order in the strange quark mass, 
the separation between the Fermi momenta is $\de p_F = M_s^2/(4\mu)$, 
so the splitting is smaller at higher densities.
Electrons are also present in weak equilibrium, with 
$\mu_e=M_s^2/(4\mu)$, so their charge density is parametrically of 
order $\mu_e^3\sim M_s^6/\mu^3\ll\mu M_s^2$, meaning that they are 
unimportant in maintaining neutrality.

In the CFL phase the situation is completely different.
All the colors and flavors pair with each other, 
locking all their Fermi momenta together at a common value 
(Fig.~\ref{fig:splitting}, right panel). This is possible as long 
as the energy cost of forcing all species to have the same Fermi 
momentum is compensated by the pairing energy that is released by 
the formation of the Cooper pairs. Working to lowest
order in $M_s^2$, we can say that parametrically the cost is
$\mu^2\de p_F^2 \sim M_s^4$, and the pairing energy is 
$\mu^2\De_{\rm CFL}^2$, so we expect CFL pairing to become
disfavored when $\De_{\rm CFL} \lesssim M_s^2/\mu$; actually
the CFL phase remains favored over the unpaired phase as long as 
$\De_{\rm CFL}> M_s^2/4\mu$ \cite{Alford:2002kj}, but already becomes
unstable against unpairing when $\De_{\rm CFL}\gtrsim M_s^2/2\mu$
\cite{Alford:2003fq,Alford:2004hz}.
Schwinger-Dyson calculations \cite{Nickel:2006kc} confirm that the CFL
tends to be favored over other phases like the 2SC phase, and
NJL model calculations
\cite{Alford:2004hz,Fukushima:2004zq,Abuki:2004zk,Blaschke:2005uj,Ruester:2005jc}
find that if the attractive interaction were strong enough to induce
a 100~MeV CFL gap when $M_s=0$
then the CFL phase would survive all the way down to the transition
to nuclear matter. 
Otherwise, there must be a transition to some other
quark matter phase: this is the ``non-CFL'' region shown schematically 
in Fig.~\ref{fig:phase}.

When the stress is small, the CFL pairing can 
bend rather than break, developing a condensate of $K^0$ mesons, 
\cite{Bedaque:2001je}.
When the stress is 
larger, however, CFL pairing becomes disfavored. A comprehensive 
survey of possible BCS pairing patterns shows that all of them suffer 
from the stress of Fermi surface splitting \cite{Rajagopal:2005dg}, 
so in the intermediate-density ``non-CFL'' region we expect more 
exotic non-BCS pairing patterns.

\section{Compact star phenomenology}
\label{sec:phenomenology}

The high density and relatively low temperature required to produce
color superconducting quark matter may be attained
in compact stars (neutron stars). This opens up the possibility of
using astronomical observations to obtain
data on the phase diagram of quark matter,
although it must be admitted that a neutron star is not an ideal
laboratory. Most of them are thousands of light-years from earth,
and this limits the features that we can observe.
Even so, there is an ongoing effort to develop signatures for the
presence of quark matter in neutron stars (for a longer review see
\cite{Alford:2007xm}). Many of these
exploit the expected color superconductivity of quark matter,
which has a a profound effect
on transport properties such as mean free paths,
conductivities and viscosities. In this section we give a brief summary,
concentrating on neutron stars with quark matter cores,
(``hybrid stars''). Pure quark matter stars (``strange stars'') only
exist if quark matter is more stable than nuclear matter 
even at zero pressure, and we will not discuss that possibility.

\subsection{Quark matter and the mass-radius relation}
In principle one might think that color superconductivity should
affect the mass-radius relation for neutron stars with quark matter cores,
(``hybrid stars'') since it affects the equation of state (EoS)
at order $(\Delta/\mu)^2$ \cite{Alford:2002rj,Lugones:2002va}.
However, other parameters such as the 
effective strange quark mass can have similar effects on the EoS,
so it is hard to distinguish color-superconducting
quark matter from unpaired quark matter  using the $M(R)$ curve.

Actually, the $M(R)$ curve does not clearly tell us whether there
is {\em any} kind of quark matter in the star. Some authors
have relied on the idea that quark matter is ``soft'', which would
mean that a hybrid star has a low maximum mass, and so
finding a neutron star with a mass of order $1.8~M_\odot$ or higher
would rule out the presence of quark matter in its core
(e.g.~\cite{Ozel:2006bv}). This is true for matter consisting of
free quarks, but when one includes reasonable estimates of
strong interaction corrections the quark matter EoS becomes
considerably stiffer. Hybrid stars can then have masses up to
$2~M_\odot$, and their $M(R)$
curves become almost indistinguishable from those predicted by
commonly-used models of nuclear matter \cite{Alford:2004pf,Alford:2006vz}.

\subsection{$r$-mode spindown}
\begin{figure}
\parbox{0.45\hsize}{
%\hrule % to gauge size of parbox
\begin{center}
\small \underline{\underline{Nuclear matter}}
\end{center} \vspace{-1ex}
\includegraphics[width=\hsize]{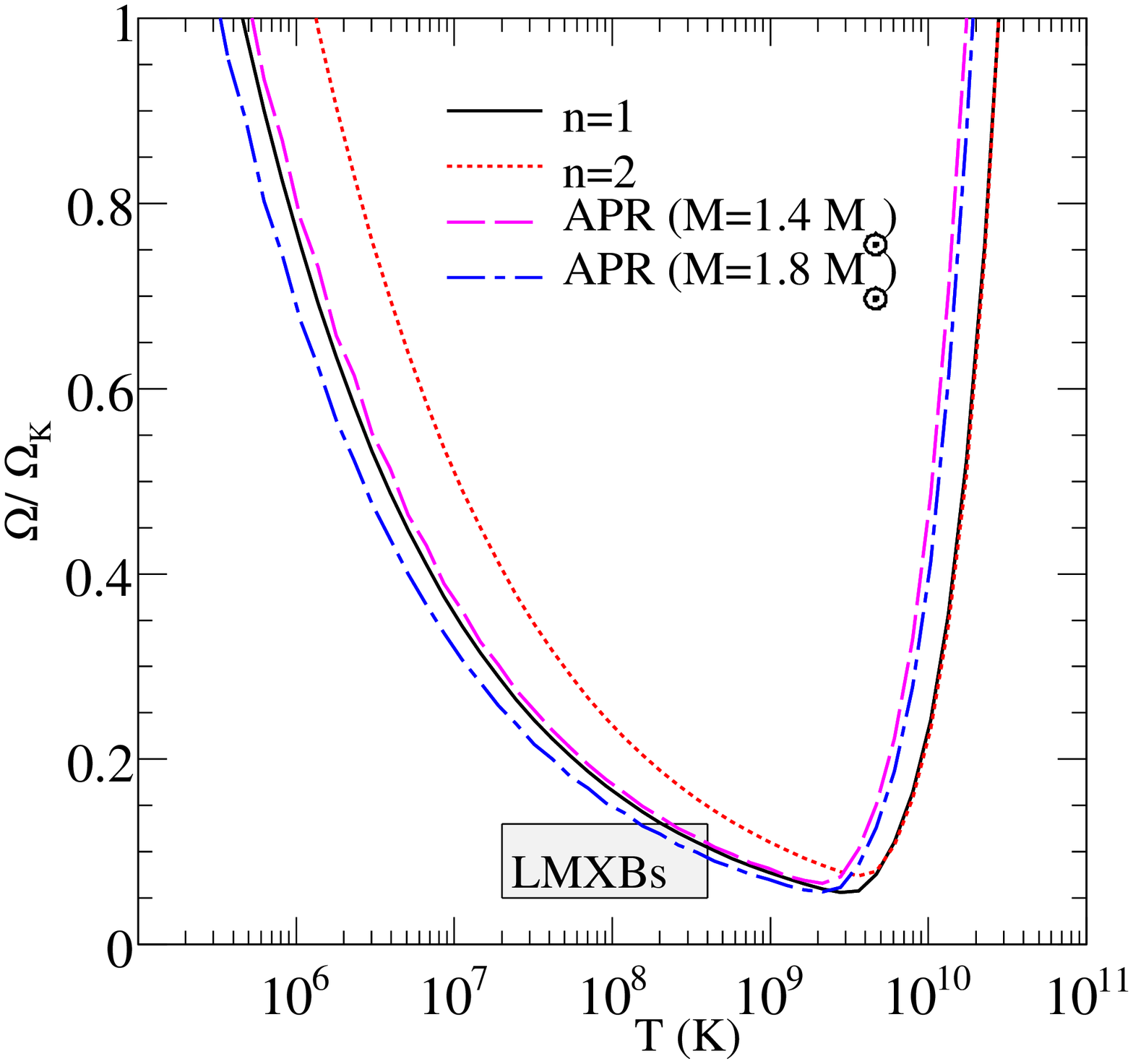}
}\hspace{0.05\hsize}\parbox{0.45\hsize}{
%\hrule % to gauge size of parbox
\begin{center}\small \underline{\underline{Quark matter}}
\end{center} \vspace{-1ex}
\includegraphics[width=\hsize]{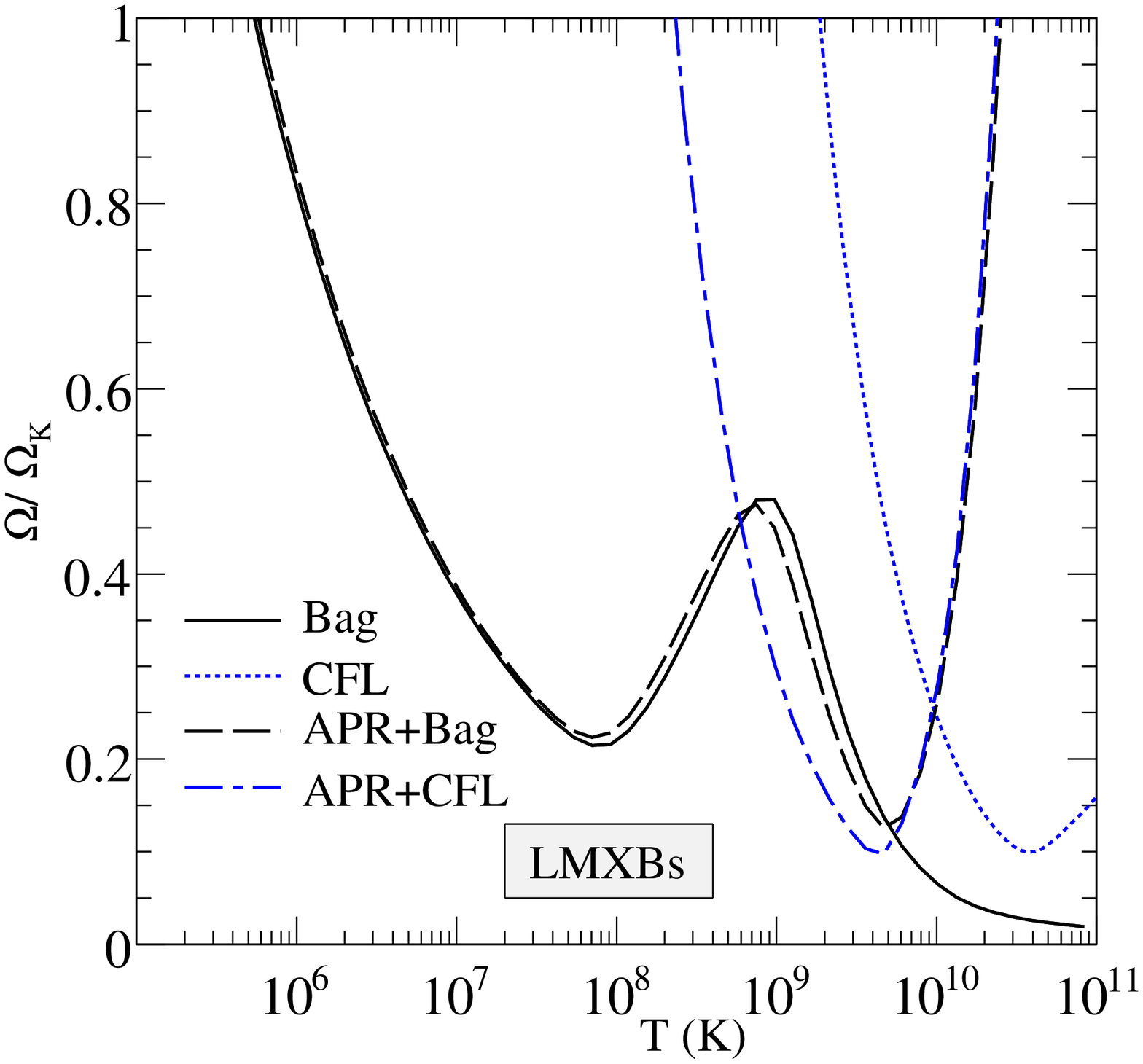}
}
\caption{
Forbidden regions (areas above the curves) of spin frequency
$\Om$ and temperature $T$ for neutron stars, predicted by
various models of their interiors. Left panel: models
of nuclear matter. Right panel: models of quark stars and
hybrid stars. Observed values for low-mass X-ray binaries
fall in the box marked ``LMXBs''.
For details, see Ref.~\cite{Jaikumar:2008kh}.
}
\label{fig:JRS}
\end{figure}

The $r$-mode is a bulk flow in a rotating star that, if the bulk and shear
viscosities are low enough, spontaneously arises and
radiates away energy and angular momentum
in the form of gravitational waves
\cite{Andersson:1997xt,Friedman:1997uh}.
Since viscosity is a sensitive function of temperature, this
leads to ``forbidden regions'' in the $\Om$-$T$ (spin frequency vs
temperature) plane: any star that started off in such a region would
quickly spin down via the excitation of $r$-modes, and exit the region.
Any hypothesis about the interior constitution of a neutron star
will lead to predictions of its viscosity, and hence a characteristic
forbidden region in the $\Om$-$T$ plane. This is illustrated in
Fig.~\ref{fig:JRS}, which is taken from Ref.~\cite{Jaikumar:2008kh}.
We see that the forbidden region for various models of nuclear matter
(left panel) is quite different from that for models of hybrid stars 
(right panel).
The analysis neglects potentially important features, such as
mutual friction (phonon-vortex scattering) \cite{Mannarelli:2008je} and
modification of the $r$-mode profile by the non-uniformity of the
star, but illustrates how astrophysical observations can probe
neutron star interiors.

\subsection{Quark core density discontinuity and gravitational waves}
The interface between a quark matter core and a nuclear matter
mantle could be a sharp interface with a
jump in energy density. (The alternative is a mixed
phase with a smooth density gradient, but this
only occurs if the surface tension of the interface
is less than about $40~{\rm MeV}/{\rm fm}^2 = 0.2\times (200~{\rm MeV})^3$, 
a fairly small value
compared to the relevant scales $\Lambda_{\rm QCD}\approx 200~{\rm MeV}$,
$\mu\sim 400~{\rm MeV}$ \cite{Alford:2001zr}.)
A sharp interface might modify the signature of gravitational waves emitted
during mergers and detected via observatories like LIGO, 
since those encode information about the ratio
$M/R$~\cite{Faber:2002zn}, and the star would in effect have two
radii, one for the quark core and one for the whole star.

\subsection{Crystalline pairing, gravitational waves, and pulsar glitches}
One candidate for the intermediate  ``non-CFL''
quark matter phase of Fig.~\ref{fig:phase} is the ``LOFF'' crystalline phase 
\cite{Alford:2000ze}. Current indications are that the crystal
has a much higher shear modulus ($\nu\sim 0.5$-$20~\MeV/\fm^3$)
than nuclear matter ($\nu\sim 10^{-4}$-$10^{-2}~\MeV/\fm^3$)
\cite{Mannarelli:2007bs}. One resultant signature is that the quark
matter is rigid enough to sustain a large
quadrupole moment, leading to detectable emission of
gravitational waves. The LIGO non-detection of such gravity waves
from nearby neutron stars already shows that they do not have
quark matter cores that are deformed to the maximum extent allowed by the 
estimated shear modulus \cite{Haskell:2007sh,Lin:2007rz}.

Two other relevant phenomena are glitches, in which pulsars
speed up their rotation occasionally, 
and precession. However, it is hard to come up with a mechanism that
allows for both these phenomena in the same star \cite{Link:2008aq}.
The standard glitch
mechanism involves pinning of superfluid vortices in the crust, which would
suppress precession in all stars, since they all have crusts.
Quark matter offers a way out---glitches could arise from pinning in
a crystalline quark matter core.
Then there would be two populations: heavy stars with a crystalline core
which could glitch but not precess; and lighter stars with no core
which could precess but not glitch.
To test this we need better calculations of the properties of the crystalline
phase and more detailed observations
of glitch rates and precession frequencies.

\subsection{Cooling by neutrino emission}
The cooling rate is
determined by the heat capacity and emissivity, both
of which are sensitive to the spectrum of low-energy excitations,
and hence to color superconductivity.
CFL quark matter, where all modes are gapped, has a much
smaller neutrino emissivity and heat capacity than nuclear matter, and
hence the cooling of a compact star is likely to be dominated by the
nuclear mantle rather than the CFL core 
\cite{Shovkovy:2002kv,Jaikumar:2002vg}.  
Other phases such as 2SC or LOFF give large gaps to only
some of the quarks. Their cooling would proceed quickly, then
slow down suddenly when the temperature fell below
the smallest of the small weak-channel gaps. This behavior should be
observable \cite{Grigorian:2004jq,Aguilera:2005tg,Anglani:2006br}.
There is already evidence that, 
although the cooling of many neutron stars  is
broadly consistent with the standard cooling curves, some fraction of neutron
stars cool much more quickly \cite{Page:2009fu}.
One may speculate that
lighter neutron stars cool following the standard cooling curve and are
composed of nuclear matter throughout, whereas heavier neutron stars
cool faster because they contain some form of dense matter that can
radiate neutrinos via the direct Urca process \cite{Blaschke:2006gd}.   
This could be quark
matter in one of the non-CFL color-superconducting phases,
but there are other, baryonic, possibilities.  If this speculation is correct,
then if neutron stars contain CFL cores they must be ``inner cores'',
within an outer core made of whatever is responsible for the rapid
neutrino emission.   

%% end of main text

\section*{Acknowledgements}
The author acknowledges the support of
the Offices of Nuclear Physics and High
Energy Physics of the U.S.~Department of Energy under contracts
\#DE-FG02-91ER40628,  % Wash U theory
\#DE-FG02-05ER41375. % Mark DoE

% \begin{thebibliography}{00} % do not change 
% \end{thebibliography} % do not change 

% Give it an href command
\newcommand{\href}[2]{#2}

\bibliographystyle{apsrev_MGA}  % Need to add arXiv: to numeric ones by hand
\bibliography{qm}

% To generate the .bib file, mail this tex file
% To:       slaclib2@slac.stanford.edu
% Subject:  generate bibtex
%    tex/confproc/2009_QM/qm.tex

% export the returned file as 
% tex/confproc/2009_QM/qm.bib
% edit it to strip off mailer header
% Add entries for papers that aren't in SPIRES:
% cat non_spires.bib >> qm.bib

\end{document}